\documentclass{aastex}          
\usepackage{spr-astr-addons,xspace,natbib}    

\newcommand{\Al}{$^{26}${Al}\xspace}
\newcommand{\about}{$\simeq$}

\newcommand{\Fe}{$^{60}$Fe\xspace}

\newcommand{\Msol}{M\ensuremath{_\odot}\xspace}

\begin{document}
%
\title{Radioactive isotopes in the interstellar medium}

\shorttitle{Radioactive interstellar gas}
\shortauthors{Roland Diehl}

\author{Roland Diehl\altaffilmark{1}} 

\altaffiltext{1}{Max-Planck-Institut f\"ur extraterrestrische Physik, Giessenbachstr. 1, 
  D-85741 Garching, Germany }

\begin{abstract}
Radioactive components of the interstellar medium provide an entirely-different and
new aspect to the studies of the interstellar medium. Injected from sources of nucleosynthesis, unstable nuclei decay along their trajectories. Measurements can occur through characteristic gamma rays that are emitted with the decay, or  in cosmic material samples 
through abundances of parent and daughter isotopes as they change with decay. 
The dynamics and material flows within interstellar medium are thus accessible to measurement, making use of the intrinsic clock that radioactive decay provides. We describe how measurements of radioactive decay have  obtained a break-through in 
 studies of the interstellar medium, after first summarizing the characteristics of radioactivity and the sources of unstable nuclei. 
\end{abstract}


%
\section{Introduction}\label{sec:intro}
\subsection{Modern astronomy and radioactive nuclei}

\begin{figure}[t]
\centering 
\includegraphics[width=\columnwidth]{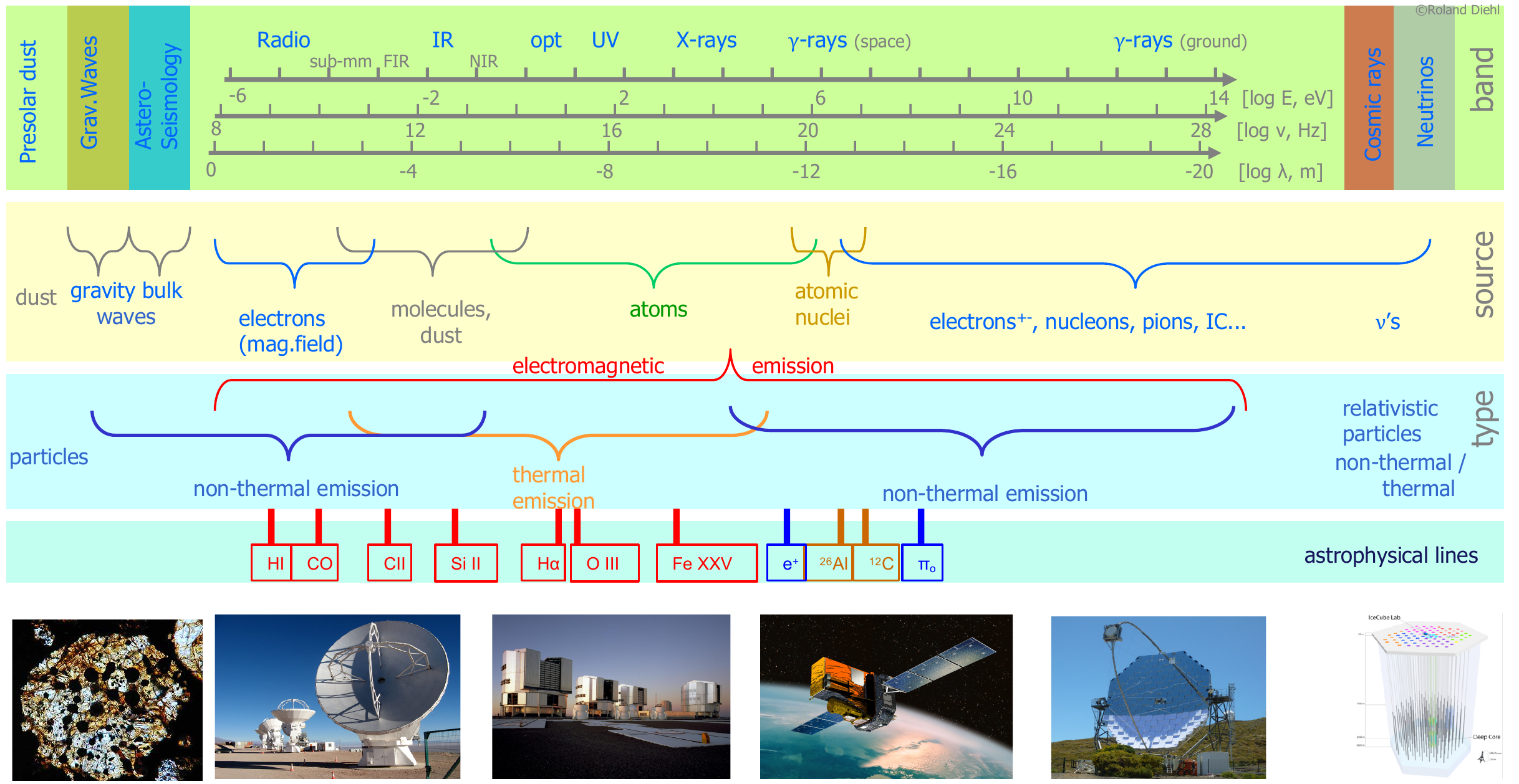}
\caption{The electromagnetic spectrum, from radio wavelengths to gamma-rays, covers more than 20 orders of magnitude, and includes signatures of different physical processes, as labelled. Specialised telescopes address mostly narrow astronomical windows, some from ground (optical, radio), and many on space platforms that have boosted the broad coverage of the different domains of the electromagnetic spectrum. Non-radiation based astronomies have been developed and complement these, such as measuring neutrinos and gravitational waves, and dust particles captured within the solar system by space satellites and on Earth and Moon.}
\label{fig:astronomyMethods}
\end{figure}

A part of the gas in the interstellar medium is radioactive. Injected from sources of nucleosynthesis, radioactive newly-formed nuclei add special opportunities for studying the dynamic and complex interstellar  medium. The radioactive decay imprints a clock into the abundance of these nuclei, as they decay from parent into stable daughter nuclei. 
The table of known isotopes holds about 40--50 radioactive isotopes with lifetimes that, on one hand, exceed the first 100 years where injection physics may dominate, and, on the other hand, are shorter than the age of the Galaxy. This makes such isotopes suitable, in principle, for a new type of study of the properties of the interstellar medium itself, using radioactivity as a property.

\begin{figure}[t]
\centering 
\includegraphics[width=\columnwidth]{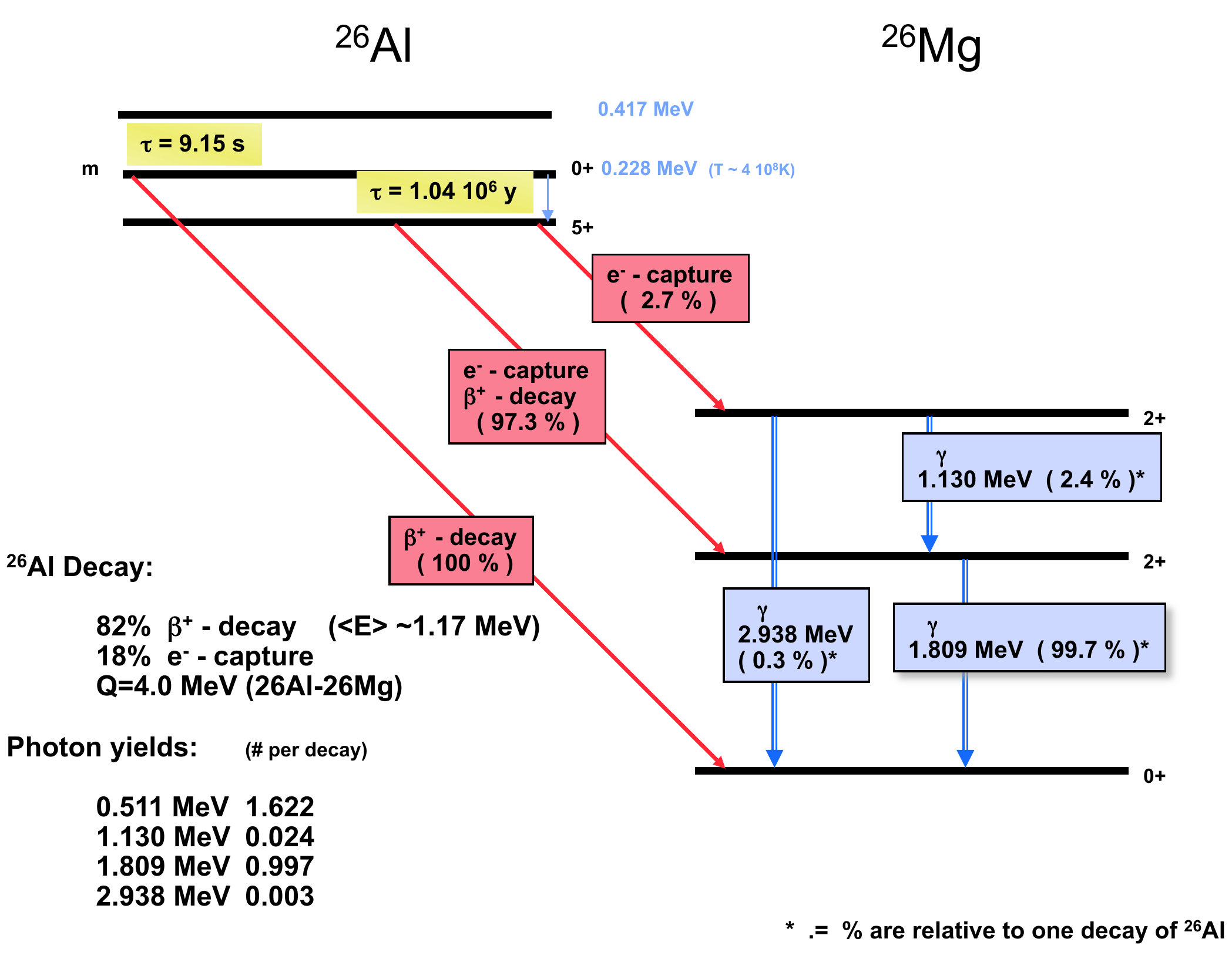}
\caption{Radioactivity is illustrated through the nuclear level and decay scheme of $^{26}$Al decay (simplified): Gamma~rays are emitted upon de-excitation of nuclear states in the daughter nucleus. For decay of $^{26}$Al, this includes gamma-rays from annihilation of the positrons produced in $\beta^+$~decay.}
\label{fig:26AlDecay}
\end{figure}
\begin{figure}[t]
\centering 
\includegraphics[width=0.8\columnwidth]{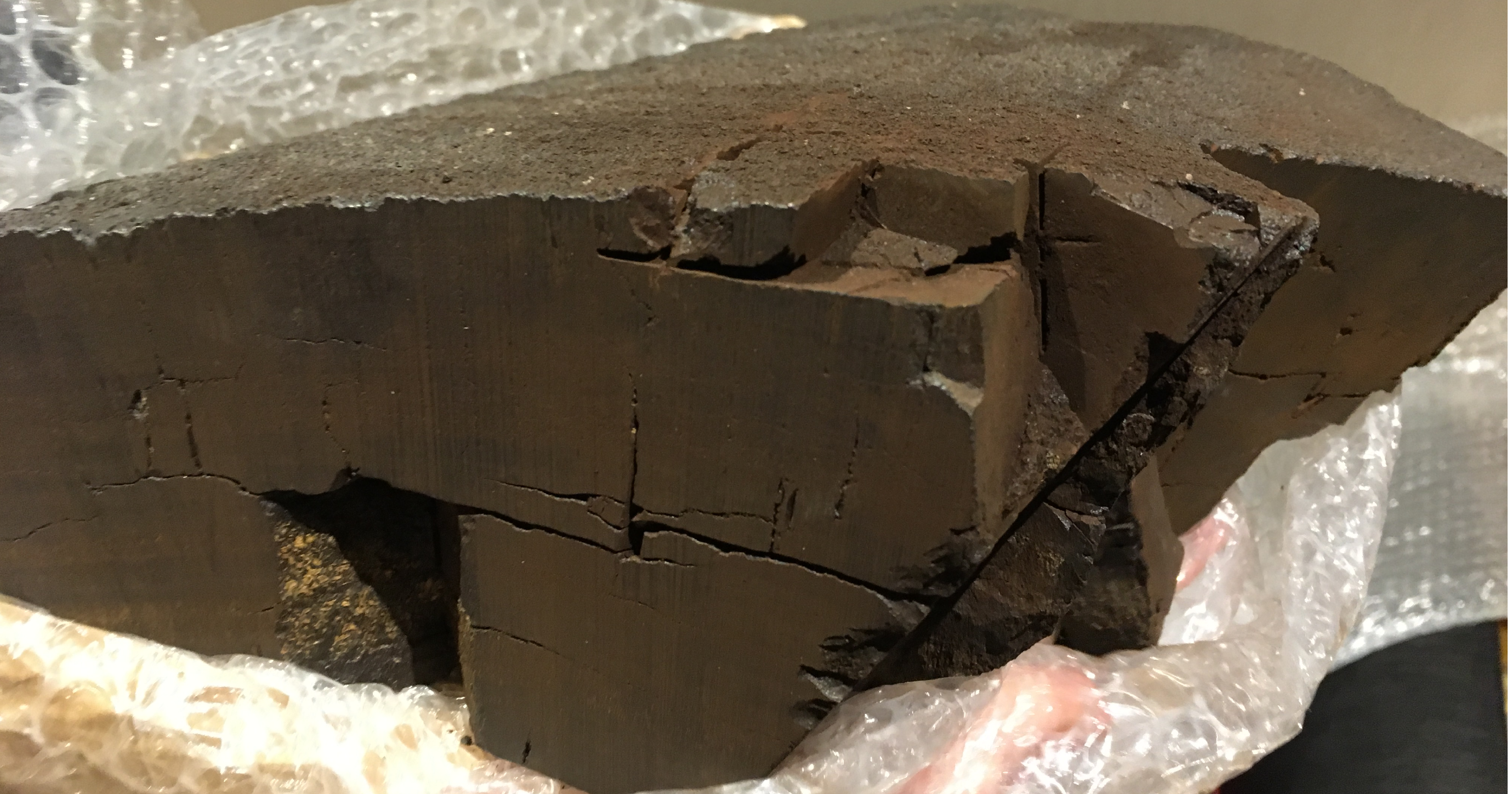}
\caption{This sample of a slowly-growing Manganese crust was retrieved from the deep pacific oceanfloor, remote from contaminations of debris of civilisation (photo courtesy Anton Wallner).}
\label{fig_oceancrust}
\end{figure}

Astronomical measurements of radioactivity use two main methods: Characteristic nuclear emission lines measured with gamma-ray telescopes, and isotopic abundances in samples of cosmic matter captured within our solar system.  
Both of these methods are rather new and not-so familiar disciplines of astronomy, as illustrated in Fig.~\ref{fig:astronomyMethods}.

Radioactive decay of an unstable isotope is often accompanied by emission of characteristic radiation, gamma-ray line emission with energies that are specific to the nuclear levels of the daughter nucleus. 
Fig.~\ref{fig:26AlDecay} shows the nuclear level and decay scheme for one example, the $^{26}$Al isotope.
This isotope has a lifetime of 1.04~Myrs, and decays mostly by $\beta^+$-decay into an excited state of $^{26}$Mg, which emits its excitation energy through characteristic gamma-ray line emission with an energy of 1808.63~keV. 
$^{26}$Al was the first unstable isotope which was detected to decay in the interstellar medium \citep{Mahoney:1982}. Its lifetime is shorter than typical times of stellar evolution, and much shorter than the age of the Galaxy itself or older stars herein. So it must have been produced recently and ejected from young, massive, and shortlived stars and possibly their terminal supernovae \citep[see][for reviews]{Prantzos:1996,Diehl:2021}.

Cosmic material had been the seed for all bodies of our solar system, including Earth, Moon, and meteorites. Today, cosmic dust and nuclei of high-energy cosmic rays reach instruments that have been launched into near-earth space. 
Analysing cosmic-material probes in terrestrial laboratories has provided rich detail about the full variety of isotopes and their abundances, both for cosmic rays and for dust particles in their great variety of mineral phases \citep{Zinner:2008}.
Cosmic gas and dust that reaches our planet Earth from cosmic material flows may end up in sediments on Earth and moon.
These minute admixtures of cosmic origins 
only could be found in a very sensitive method called \emph{accelerator mass spectrometry} \citep{Kutschera:1990}.
Other methods to analyse samples of cosmic materials include the ion-beam sputtering of meteoritic samples within \emph{secondary-ion mass spectrometry, NanoSIMS} instruments \citep{Hoppe:2004} to determine precision isotopic abundances in small presolar grains, and thus measure characteristic daughters from radioactive decays.
Through experimental efforts such as these, one may, therefore, consider the analysis of samples of cosmic materials also as an \emph{astronomy}, though providing information about astrophysical processes and sources in different ways.
Clearly, these methods are subject to a fundamental constraint of \emph{how} the material sample was obtained, or formed in the first place.
Upon closer examination, we note that also electromagnetic observations that have been used for species abundance determinations involve constraints of this type; e.g., a species must be radiation-coupled, thus in case of atomic line emission be not fully-ionised and excited by, e.g., collisions.

\subsection{Radioactivity and astrophysical studies} 

\begin{figure}
\centering 
\includegraphics[width=\columnwidth]{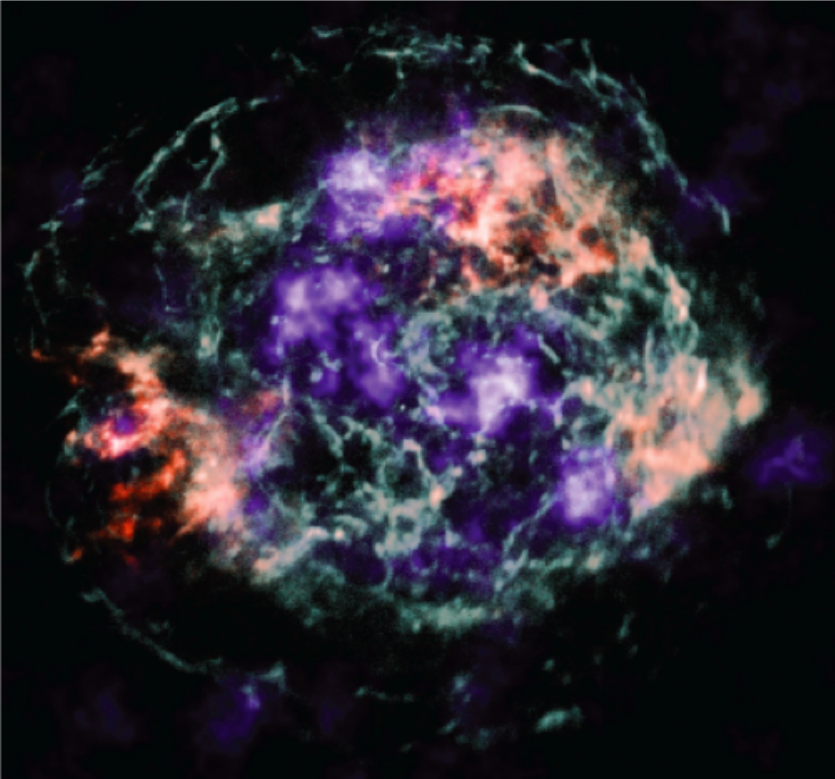}
\caption{The image of the Cas A supernova remnant demonstrates how radioactivity complements our view: Characteristic lines from $^{44}$Ti decay (blue) reveal the location of inner ejecta, while X-ray line emissions from iron (red) and silicon (green) atoms, that also are emitted from those inner ejecta, show a somewhat different brightness distribution, due to ionization emphasising the parts of ejecta that have been shocked within the remnant. From \citet{Grefenstette:2014}.}
\label{fig:CasAimage}
\end{figure}

The release of nuclear binding energy, as in radioactive decay, plays a prominent role for the appearance of many sources in astronomy throughout the universe.
Stars shine, because nuclear reactions in the inner regions fuse lighter to heavier nuclei, which are more-tightly bound \citep{Eddington:1919}. 
Successive burning stages during the evolution of a star correspond to different nuclear fuels, each of which temporarily slows down the gravitational collapse of the star (Fig.~\ref{fig:stellar-burning}).
In supernova explosions, large amounts of radioactive $^{56}$Ni are produced: Of order 0.1~\Msol are typically produced in core-collapse supernovae, as we know from SN1987A (0.07~\Msol)\citep{Arnett:1989a,Fransson:2002,McCray:2016}.
SN1987A was the first core-collapse supernova where gamma rays directly attributed to the radioactive energy input from $^{56}$Ni decay could be seen: The Solar Maximum Mission and its Gramma-Ray Spectrometer instrument showed the characteristic lines from decay of $^{56}$Co at 847 and 1238 keV, respectively \citep{Matz:1988}. 
Thermonuclear supernovae (type Ia) produce even more, typically 0.5~\Msol, as we know from e.g. SN2014J \citep{Scalzo:2014,Seitenzahl:2017,Diehl:2015}.
The gamma-rays and positrons emitted from radioactive decay of $^{56}$Ni energise the ejecta and envelope of the supernova from inside, and the scattered and re-processed radioactive energy is thus responsible for the light that appears at the \emph{photosphere} and makes supernovae bright sources. 
Although this photospheric emission is key to many astrophysical applications,  the radioactivity also directly provides a diagnostic of the explosions themselves.
The radioactive energy injection of $^{56}$Ni decay to power light from thermonuclear supernovae could be measured directly for the first time with INTEGRAL's spectrometer SPI \citep{Diehl:2014,Diehl:2015,Churazov:2014}. This is a valuable complement, as it bypasses the uncertainties and complexities of radiation transport within the supernova, that downscatters MeV radiation by many orders of magnitude into optical light. Its diagnostic power had been emphasised by modeling work \citep[e.g.][]{Summa:2013} long before SN2014J occurred.
The radioactive afterglow of supernova remnant Cas A in $^{44}$Ti line emissions \citep{Grefenstette:2014} showed us the morphology of young ejecta, as they still are forming a remnant that also is bright in X-rays and radio emissions (Fig.~\ref{fig:CasAimage}). $^{44}$Ti has a radioactive lifetime of 89 years, and thus had been helpful to study this remnant with about 350~years of age \citep{Vink:2000,Vink:2012}.
Also fluorescent X-ray line emission at keV energies may be emitted from radioactive material, if decay occurs through electron capture and leads to such X-ray emission as the atomic-shell vacancy is replenished \citep{Seitenzahl:2015}; this may have added to the energisation of the late SN1987A light curve, but direct detections remain ambiguous \citep{Borkowski:2010}.

X-ray and radio continuum emissions of supernova remnants fade within thousands of years. 
The later supernova remnant phases before they dissolve and merge with ambient gas can be studied, if surrounding gas is dense enough, and emission re-brightens upon the interaction of the supernova shock with the surrounding interstellar gas, and the reverse shock also illuminates the remnant itself from the inside.

Beyond and at a later stage, longer-lived radioisotopes among the supernova ejecta  trace what happens as ejecta merge into the ambient interstellar medium; this will be discussed in more detail in Section~\ref{sec:astroApplications} below.

Radioactive decay is independent of the thermodynamic state and condition of the gas. Therefore, the imprinted clock is independent of these, and just a property of the specific nuclide. Thus, astronomical measurements of radioactive isotopes allow the study of propagation effects within the interstellar medium, using radioactivity tracing\footnote{We are familiar with measurements tracing radioactivity in medical applications}.

Beyond studies in electromagnetic radiation, radioactivity has been exploited in samples of cosmic materials for astrophysical studies.
Cosmic dust particles, called \emph{stardust}, have been incorporated into meteorites that formed in the early solar system, and can be found therein. Their characteristic isotopic abundances can be exploited towards constraints on the formation sites of these dust particles. Grains attributed to AGB stars, but also to nova and supernova origins, have been discussed \citep{Zinner:1998,Clayton:2004,Zinner:2008}.
Cosmic rays captured with suitable instruments on space satellites from nearby interplanetary space have been found to include several radioactive isotopes, including $^{14}$C, $^{36}$Cl, $^{26}$Al, $^{10}$Be, $^{59}$Ni, and others. 
Data from 
satellites \citep{Israel:2005,Israel:2018}, high-altitude balloons  \citep{Walsh:2019}, and the space station \citep{Adriani:2020} have been used to constrain the propagation, specifically the  path lengths traversed,  for cosmic rays, as their spallation reactions on heavier nuclei such as Fe have created these radioisotopes. 
For the case of  $^{59}$Ni, its detection constrains the time between ejection from its nucleosynthesis source and the acceleration to cosmic-ray energies, because $^{59}$Ni only decays through electron capture, and thus remains stable once fully ionised in cosmic rays after acceleration \citep{Mewaldt:2001,Wiedenbeck:2001a}. 
Other dust and gas particles of cosmic origins also continuously reach our planet Earth and the moon, and are deposited in minute amounts as sediments. They had first been detected   in a sample of an ocean crust from the deep pacific ocean floor, see Fig.~\ref{fig_oceancrust}, where slow sedimentation of materials has been found to include cosmic radioactivity in the form of $^{60}$Fe isotopes \citep{Knie:1999,Wallner:2016}. More details and astrophysical lessons will be discussed in Sect. 3.2. below.

The context of this review is approached here from a viewpoint of nuclear gamma-ray spectroscopy.  Nevertheless, we will also address other astronomical tools towards interstellar radioactivity, at least briefly.
In this paper, we first present an overview of the different sources that are believed to be responsible for radioactive materials within the interstellar medium.
Thereafter, we discuss the two main different astrophysical areas addressed by studies with radioactive tracers, (1) the transport of new nuclei from its sources towards mixing with the ambient interstellar gas, and (2) the use of radioactive tracers to learn about metallicity enrichment within galactic gas in general, within the cycle of matter including successive stellar generations.

\section{Sources of radioactive nuclei}\label{sec:sources}

\begin{figure}
\centering
  \includegraphics[width=\linewidth]{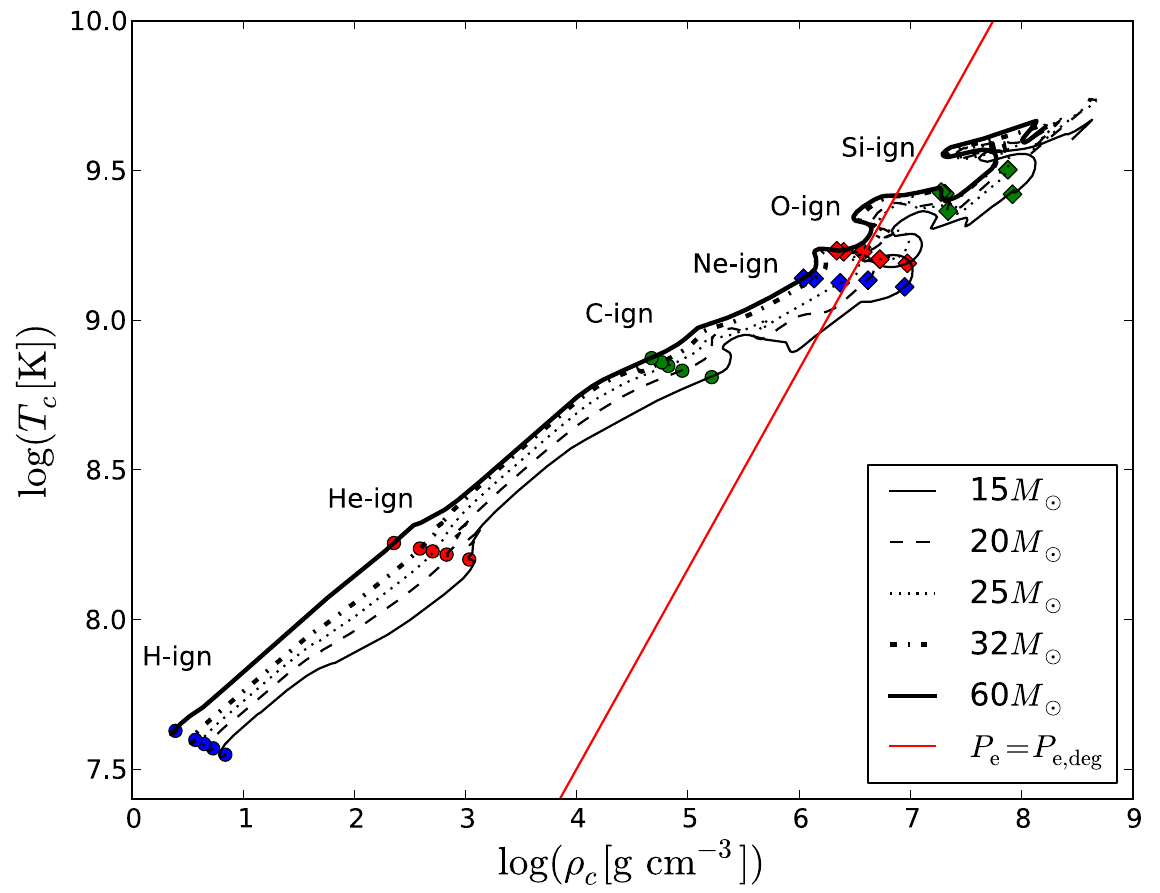}
   \caption{The track of stellar evolution for massive stars, including the different nuclear-burning stages. Shown is central density versus central temperature, for different evolutionary-track calculations using the NUGRID software. The dashed line indicates the transition into degenerated gas in the core. From \citet{Pignatari:2016}.}
  \label{fig:stellar-burning}
\end{figure}

Radioactive substances are created by nuclear reactions mainly within stars and their explosions, and to some extent also within the interstellar medium.
In stars, as densities and temperatures exceed the threshold for nuclear reactions within the stellar core. The trajectory in temperature and density for nuclear burning within massive stars is shown in Fig.~\ref{fig:stellar-burning}. Within stellar explosions, densities within the inner supernova approach values for nuclear matter, and temperatures exceed GK. Under such conditions, matter experiences what is called \emph{nuclear statistical equilibrium}, and the nucleons available from the local composition of matter, i.e. protons and neutrons, mostly with a mild excess of the population of neutrons, is processed as guided by the binding energy of nuclei that may form in this region of phase space. 
For example, if the fuel consists of nuclei of equal proton and neutron numbers, such as $^4$He or $^{28}$Si, the main product would be $^{56}$Ni; generally, nuclear binding energy is maximised within the realm of iron-group nuclei. 
Such density and temperature is only realised in explosive environments, and possibly near highly compact and dynamic accretion disks that are in the process of forming a black hole. 
Thermonuclear supernova explosions that are launched from compact white dwarf stars at densities near 10$^{10}$g~cm$^{-3}$ clearly satisfy such conditions, and hence their main burning products are nuclei of the iron group, with a substantial amount of radioactive $^{56}$Ni of about 0.5~\Msol \citep[see][for a review]{Seitenzahl:2017}. 
Black-hole accretion disks will also heat up matter to nuclear burning temperatures at their outer radius, with corresponding nucleosynthesis towards equilibrium burning \citep{Surman:2006}. It remains to be seen if parts of such matter could possibly be ejected, owing to the dynamical extremes that may occur as a compact star merges into the black hole, or explosion kicks of a preceding core-collapse supernova that first formed a neutron star before fallback material increased its mass beyond stability.  

\begin{figure}[t]  
\centering 
\includegraphics[width=\columnwidth]{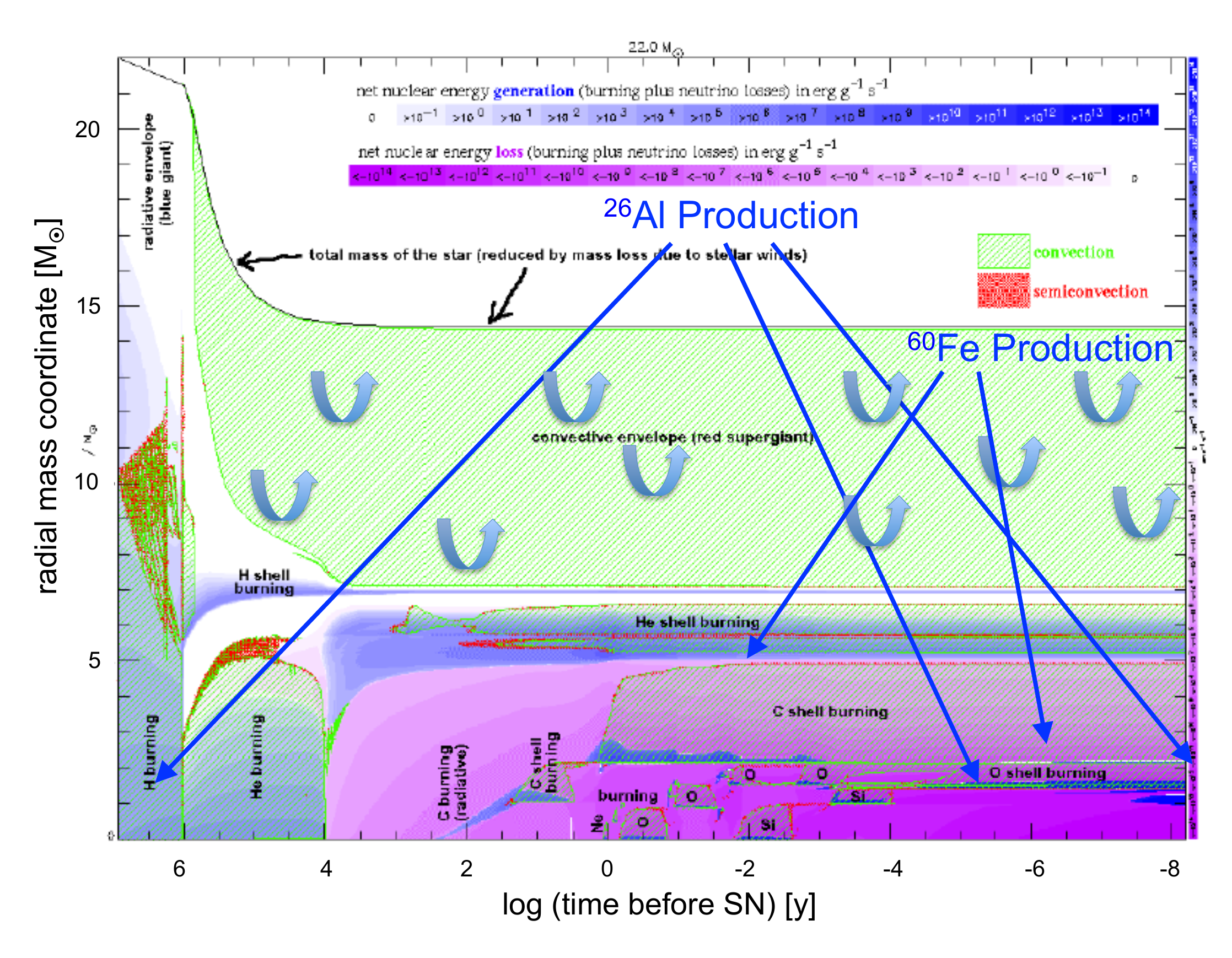}
\caption{The radioactive isotopes and gamma-ray emitters  \Al\ and \Fe\ are both produced in massive stars, albeit in different regions of the star and at different times during evolution. In the representation of a massive star's evolution of this graph (Kippenhahn diagram, courtesy A. Heger) the production regions are indicated. While hydrostatic \Al\ production early on may end up in the surface mass lost from wind, all later and more-inside productions can only be ejected with the supernova. }
\label{fig:AlFe_productionRegions} 
\end{figure}   

Within stars, the particular nuclear reactions obey a complex network of reactions, that involve also unstable nuclei produced alongside, and products vary per source type and environment. As Fig.~\ref{fig:stellar-burning} shows, we distinguish several main burning phases from the main fuels, starting with hydrogen (the \emph{main phase} of stellar evolution), then helium (the \emph{giant phase}, followed by carbon, neon, oxygen, and silicon burning states. 
The latter burning stages are much shorter than the main and giant phases, because neutrinos are produced within such burnings, and these carry away energy that otherwise could stabilise the star, causing more-rapid contraction and fuel consumption to avoid collapse. 
Once the central composition of the star reaches iron group nuclei, no further energy release can be obtained from nuclear fusion reactions, and the star collapses.
Depending on mass of the star, this can directly lead to a compact remnant such as  a black hole, thus burying all stellar matter forever.  
For stars less massive that about 25~\Msol, however, more-likely is the inversion of the collapse into a supernova explosion, as infalling matter bounces off the inner nuclear-density wall and is energized by neutrinos produced as protons are converted to neutrons to avoid Coulomb repulsion and form dense neutron-star matter.
This supernova explosion proceeds with the bounce shock wave propagating through the infalling envelope of the entire star, leading to some explosive nuclear burning.
The explosion thus ejects all ashes produced from nuclear burnings therein, i.e. the burning phases shown  in Fig.~\ref{fig:stellar-burning} and corresponding shell burnings outside the core.

Typically, several \Msol of material is ejected from core-collapse supernova explosions, including typically fractions less than 0.1~\Msol of radioactive nuclei. Much of the latter is $^{56}$Ni, which decays within 8~days to $^{56}$Co, which decays to stable $^{56}$Fe within 111~days. More-longlived radioactive isotopes within core-collapse ejecta include small amounts of $^{22}$Na and  $^{60}$Co with a radioactive lifetimes of 3.8 and 7.6~years, respectively, and some 10$^{-4}$~\Msol of $^{44}$Ti with a radioactive lifetime of 89 years.

Beyond the massive stars that have been described so far, other sources of nuclearly-processed materials exist. 
Most prominently, supernova explosions of another type, \emph{thermonuclear} supernovae called \emph{type Ia}, are important sources of interstellar radioactivity. Here, a thermonuclear runaway explosion of a white-dwarf star occurs, and disrupts the entire star, ejecting typically 1--1.4~\Msol of gas into its surroundings \citep{Seitenzahl:2017}. In these explosions, again nuclear-statistical equilibrium conditions are achieved, now for the majority of the processed and ejected material, making these objects bright sources usable for cosmological studies, through energy that is released as the typically 0.5~\Msol of radioactive $^{56}$Ni decay, energising the still-dense exploding matter so its photosphere is a bright source of UV-optical-infrared light. 
We believe that these supernovae are the result of mass transfer from a companion star onto the white dwarf in a binary system, thus driving the white dwarf towards Carbon ignition near the Chandrasekhar mass limit near 1.4~\Msol. The occurrence of a supernova of type Ia is typically delayed to occur at a time of order Gyr after the initial star formation \citep{Ruiter:2011}, unlike massive-star activity, which is prompt to within few ten Myrs. 
This also implies that typical source environments will be different: massive stars embedded in stellar clusters and/or near molecular clouds, whereas type-Ia supernovae may typically occur in regions that have been depleted in gas from massive-star feedback, if not the binary system has moved away from the star-formation site itself during its long evolutionary time.

Other sources of nucleosynthesis are nova explosions, which occur rather frequently, at a rate of about 30~y$^{-1}$ within our Galaxy \citep{Shafter:2017}. These also result from mass transfer within a binary system, but only require accumulation of a hydrogen mass critical to ignition on the white dwarf star.
This occurs in WD systems with mass transfer rates of order 10$^{-9}$~\Msol~y$^{-1}$, and ejects typically 10$^{-7}$ to 10$^{-4}$~\Msol of material into its surroundings   \citep{Jose:1998}.

More exotic and rare contributors are believed to result from binary evolution leading to neutron stars, and their collisions after loss of orbital energy to gravitational radiation \citep{Metzger:2017}. 
This scenario had been confirmed by observations spectacularly in 2017 with GW170817, as a trigger from a gravitational event and a short gamma-ray burst stimulated a rich diversity of observations of a kilonova and its afterglows.
Ejection of significant amounts of material is predicted, up to 10$^{-2}$~\Msol, that probably includes r-processed ejecta of heavy elements \citep{Wu:2016}.
While such rare events, which may also include a rare subtype of core-collapse supernovae called jet supernovae, may be important to understand the cosmic abundance evolution of rare species such as gold or uranium that are attributed to an \emph{r process nucleosynthesis}, they are not considered important for the dynamics and evolution of the interstellar medium of typical galaxies itself, due to their scarcity \citep{Belczynski:2018}.

Within the interstellar medium itself, nuclear reactions occur as energetic cosmic ray nuclei collide with nuclei of ambient gas \citep{Ramaty:1979,Tatischeff:2018b}. This \emph{spallation nucleosynthesis} produces characteristic unstable isotopes as abundantly-occurring nuclei of the interstellar medium such as carbon or iron are fragmented into lighter nuclei. 
The cosmic abundance of light elements Li-Be-B is understood to mainly result from such spallation in interstellar space, while contributions to other cosmic elemental abundances are minor \citep{Prantzos:2010}.  
The accretion disk environments discussed above would provide conditions of such energetic particle collisions, particularly as the polar plasma jet of the newly forming black hole provides relativistic particles that can propagate over kpc distances to eventually collide with colder interstellar gas clouds. Near active galactic nuclei, such a scenario has been used to predict nuclear de-excitation line fluxes \citep{Mannheim:1997}.
More plausibly, in young supernova remnants the outward moving shocks provide a setup that likely leads to particle acceleration and cosmic ray production. The cosmic rays that locally collide with ambient matter will thus incur de-excitation emission of nuclear origin, that should be measurable with next-generation gamma ray telescopes \citep{Tibolla:2011,Summa:2011} \citep[see][for a recent review]{Tatischeff:2018b}. Regions of high massive-star density also should provide such a cosmic-ray acceleration scenario. In the nearby Orion region, a first report of de-excitation line detections from $^{12}$C and $^{16}$O with the COMPTEL instrument \citep{Bloemen:1997b} could not be confirmed in later data re-analysis  \citep{Bloemen:1999}.

\begin{figure}[t]  
\centering 
\includegraphics[width=1.0\columnwidth]{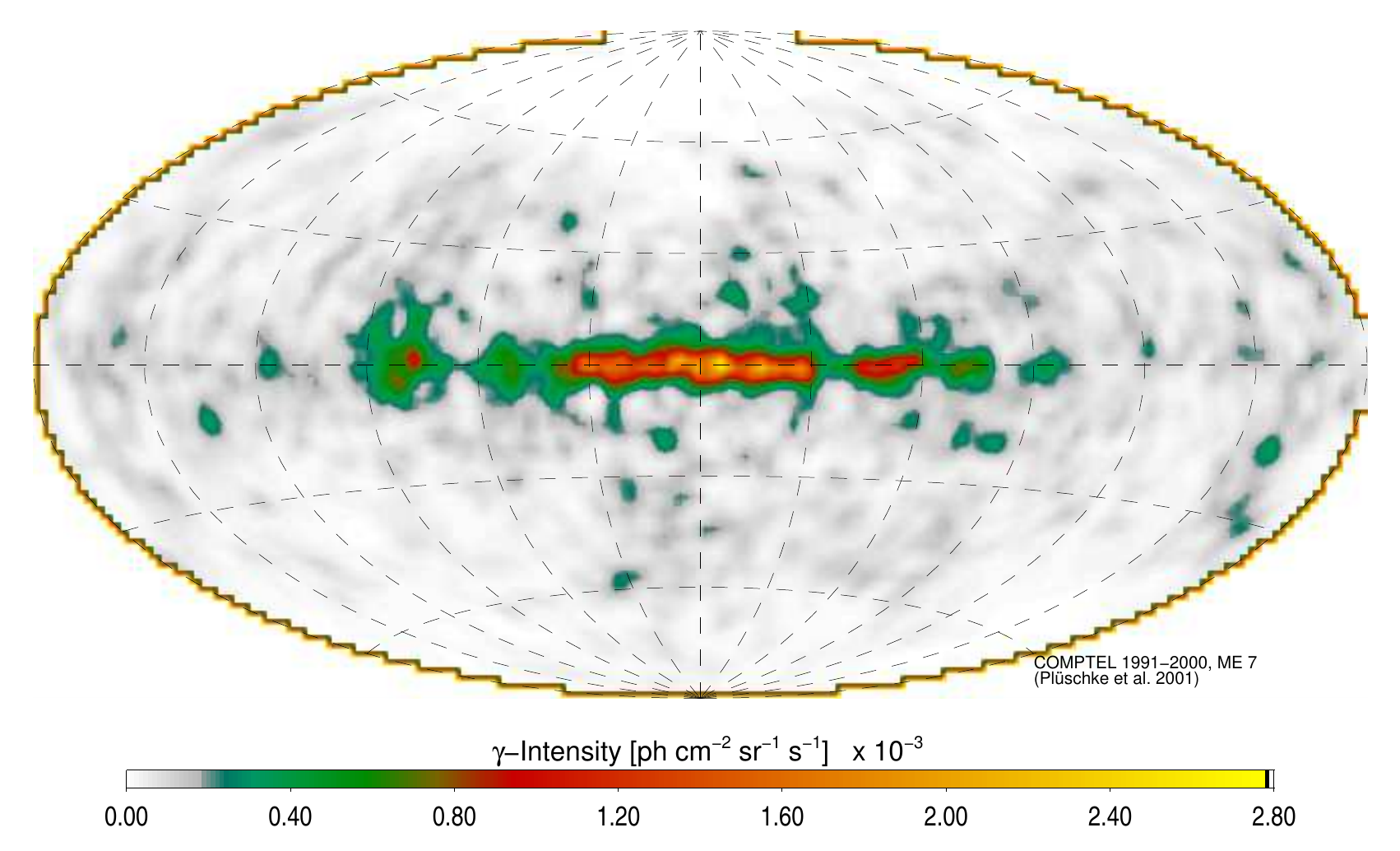}
\caption{The \Al sky as imaged with data from the COMPTEL telescope on NASA's Compton Gamma-Ray Observatory. This image \citep{Pluschke:2001c} was obtained from measurements taken 1991--2000, and using a maximum-entropy regularization together with likelihood to iteratively fit a best image to the measured photons. }
\label{fig:Al-map} 
\end{figure}   

\begin{figure}[t]  
\centering 
\includegraphics[width=0.8\columnwidth]{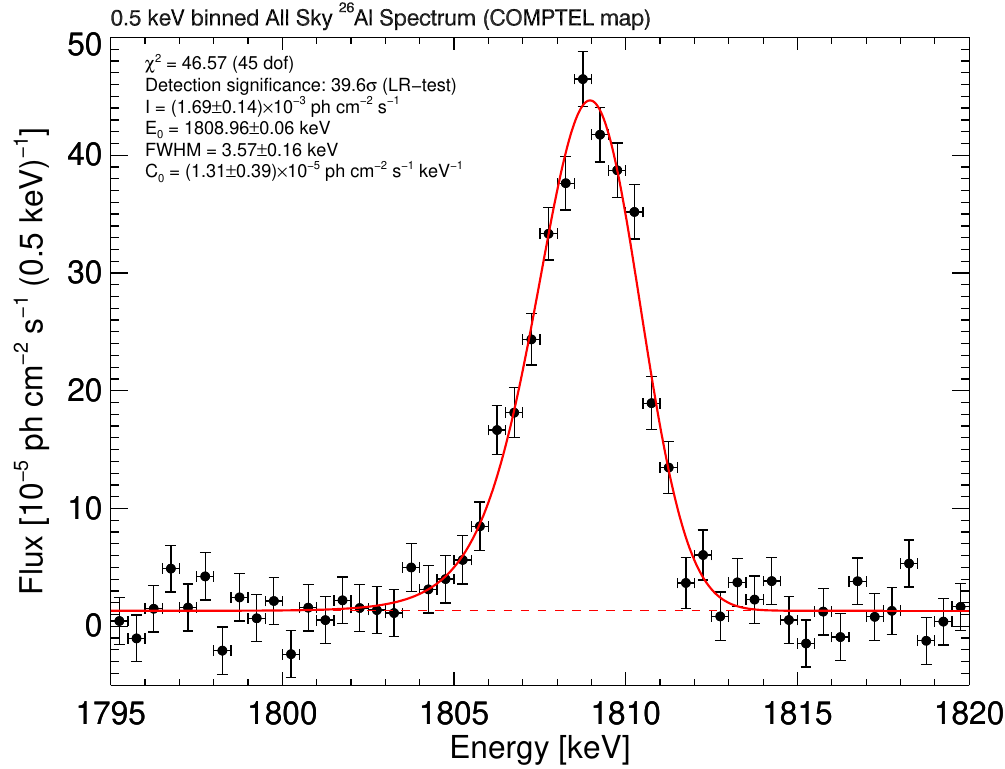}
\caption{The \Al\ line as seen with INTEGRAL's high-resolution spectrometer SPI and 13 years of measurements integrated \citep{Siegert:2017}. }
\label{fig:almap-spec} 
\end{figure}   
\section{Applications of radioactivity: astrophysical studies of the interstellar medium}\label{sec:astrophysics}
\label{sec:astroApplications}

The  presence of radioactive material within the interstellar medium is fortunate: (i) it traces current and recent nucleosynthesis, and (ii) it enables new types of investigations on the flow and the dynamics within the interstellar medium. 

Unstable nuclei are particularly useful for the study of interstellar-medium astrophysics when their radioactive lifetimes exceed the complexities of the mass ejection from nucleosynthesis sources itself. Such complexities include  processes that are partly specific to each type of source or even its particular surroundings, at least within the first few years to maybe decades. Examples are stellar winds or supernovae that interact with circumstellar matter that had been expelled before, such as superluminal supernovae. But also phenomena in the appearances of supernova remnants are driven to a large extent by the collision of the expanding explosion with ambient matter.
Here we aim to address aspects of the interstellar medium itself.

We first discuss  how radioactivity can be used to trace and study interstellar-medium properties  more generally and on the larger (multi-)~kpc scale in a galaxy. This includes complementing diagnostics of interstellar-medium parameters, and in particular the subject of compositional evolution of interstellar gas, otherwise known as the topic of \emph{chemical evolution}.
Thereafter, we discuss the local environments of the sources of radioactivity on a scale of typically 100~pc, and how the activity of those sources may affect the interstellar medium itself in their surroundings in turn; this process is called \emph{feedback}, as it shapes the star formation process that may follow the injection of fresh material into the interstellar medium. We thus arrive at nearby special examples, the Orion and Scorpius-Centaurus regions, and the neighbourhood our own solar system.

\subsection{Interstellar-medium characteristics at large}\label{sec:largescale}
Among radioactive isotopes that can be measured in interstellar space,  the gamma-ray emitters $^{26}$Al and $^{60}$Fe are most important, as they have lifetimes of Myrs.
Thus they can be observed long after newly-synthesised nuclei have been ejected into surroundings of sources. 
And, moreover, the typical nucleosynthesis injection rates (see Sect.~\ref{sec:sources}) are high enough even in entire star-forming regions to not be dependent on few individual sources. 
Their emission arises from cumulative contributions of sources of stellar nucleosynthesis over their radioactive lifetime.
Therefore, they provide a most-direct measure of such stellar nucleosynthesis activity on larger scales, from star-forming regions to the entire Galaxy.

It is \emph{most-direct}, as the emission arises from radioactive decay, a local property of a nuclear species that is independent from environmental conditions that would be characterised by thermodynamic parameters such as temperature, and also independent of density and of degrees of ionisation.
By comparison, observables such as counting of stars in the optical, measuring the temperature of dust in the infrared, and counting supernova remnants in radio emission, are less direct, as occultation, dust presence and size distributions, and ambient-medium densities, respectively, lead to systematic distortions that are difficult to assess, for a proper interpretation of measurement.

$^{26}$Al decays within 1.04~My, while $^{60}$Fe is somewhat longer-lived, with a radioactive lifetime of 3.4~My. 
\Al~ decay is directly observed  in interstellar space through its characteristic $\gamma$~rays, which have ab energy of 1808.65~keV. 
The pioneering observation of its existence and decay witjin the interstellar medium was made with the HEAO-C satellite of the US NASA space science program, as a proof of currently-ongoing nucleosynthesis in our Galaxy \citep{Mahoney:1982}. 
The COMPTEL sky survey made with the Compton Gamma-ray Observatory satellite (1991-2000) provided a first all-sky image in the \Al~ $\gamma$-ray line \citep{Diehl:1995b}.
The image showed structured \Al emission, extended along the plane of the Galaxy \citep{Pluschke:2001c,Diehl:1995b} (see Fig.~\ref{fig:Al-map}). 
This is in broad agreement with earlier expectations of \Al~ being produced throughout the Galaxy, and mostly from massive stars and their supernovae.
The similarity in spatial distribution of this emission and the inferred locations of massive and young stars within the Milky Way Galaxy led to the interpretation that these were the dominating sources of \Al\ \citep{Prantzos:1996a}, although novae and AGB stars also had been advertised as promising contributors.
It is not always straightforward, though, to interpret measurements of $\gamma$-ray line emission. 
As Fig.~\ref{fig:Al-map} shows, the inherent blurring of the measurement by the response function of the $\gamma$-ray telescopes, and their inherently-high background, require sophisticated deconvolution and fitting methods to obtain the astrophysical result. Bayesian methods need to be applied, and simple methods of subtracting background and inverting data with the instrument response matrix are not feasible. Nevertheless, the sensitivity of current $\gamma$-ray surveys penetrates to intensities as low as 10$^{-5}$~ph~cm$^{-1}$s$^{-1}$ or below. Typical exposure times are at least a month, and correspondingly longer for deeper exposures. 

A quantitative interpretation of the measured flux of \Al emission in terms of stellar nucleosynthesis activity is, however, still dependent on a fundamental difficulty in astronomical observation: Apparent brightness of a source on the sky needs to be related to the distance (and age) of the source.
The distance of sources cannot be determined from the measured gamma rays, it has to be inferred from the observed morphology of the emission, using plausible assumptions.
In Fig.~\ref{fig:Al-map}, we see that the plane of the Galaxy appears, with a brighter ridge of emission in the directions towards the central regions of the Galaxy. 

\begin{figure}[t]  
\centering 
\includegraphics[width=0.6\columnwidth]{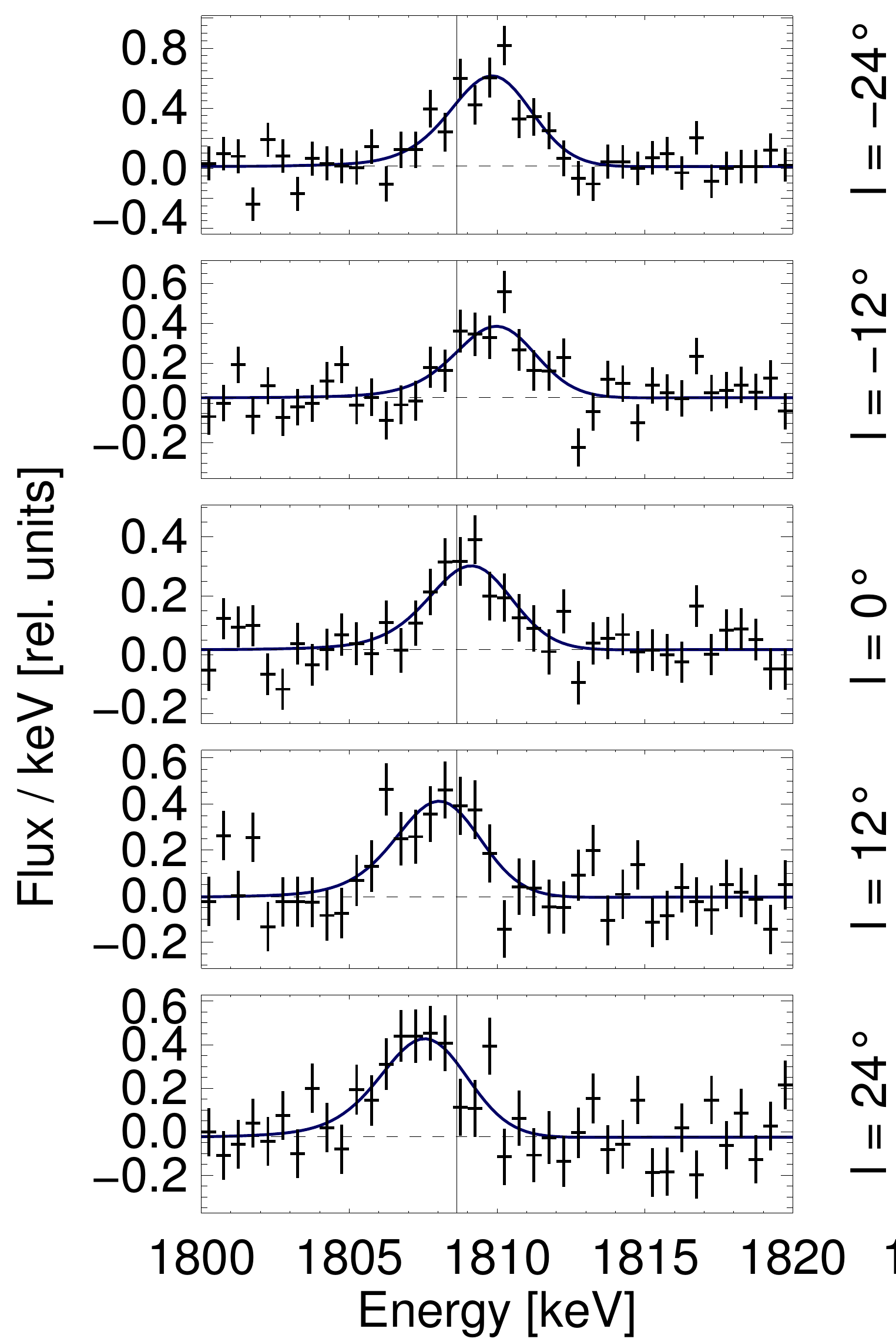}
\caption{The \Al\ line as seen towards different directions (in Galactic longitude) with INTEGRAL's high-resolution spectrometer SPI. This demonstrates kinematic line shifts from the Doppler effect, due to large-scale Galactic rotation \citep{Kretschmer:2013}.  }
\label{fig:al_longitudes} 
\end{figure}   

\begin{figure}[t]  
\centering 
\includegraphics[width=\columnwidth]{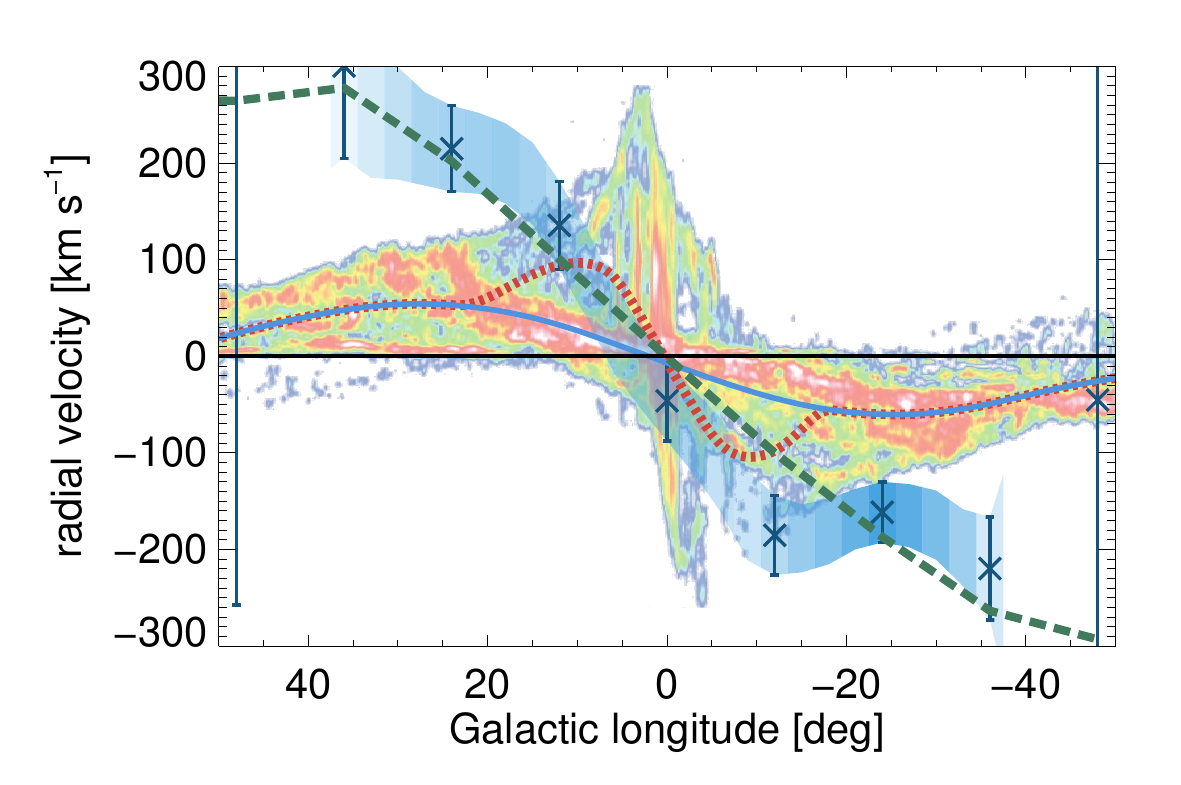}
\caption{The line-of-sight velocity shifts seen in the \Al\ line versus Galactic longitude, compared to measurements for molecular gas \citep{Kretschmer:2013}. }
\label{fig:al_long-velocity} 
\end{figure}

A second, supporting argument that the emission observed from \Al is of large-scale, galactic, origins is derived from the observed signature of large-scale Galactic rotation. 
ESA's INTEGRAL space observatory,  launched in 2002 and expected to operate till 2029, carries a Ge-detector based spectrometer called SPI.
This instrument is  suitable for high-quality spectroscopic data, with a resolution of 3~keV at the energy of the \Al line (1809~keV). This spectral resolution corresponds to a Doppler velocity shift resolution of about 100 km~s$^{-1}$ for bright source regions \citep{Kretschmer:2013}.
COMPTEL was a first-generation gamma-ray telescope; it's scintillation detectors, however, lacked the spectral resolution required, for example, for line identification and spectroscopic studies towards dynamics of the interstellar  $^{26}$Al: it had \about~200~keV instrumental resolution, compared to \about~3~keV for Ge detectors, at the energy of the \Al~ line. This had been decisive for the pioneering HEAO-C measurement to convince scientists that they had indeed detected the line from the decay of the \Al isotope, rather than any other nuclide. 
A 1995 balloon experiment (named GRIS) that carried high-resolution Ge detectors had provided an indication that the \Al~ line was significantly broadened to 6.4~keV \citep{Naya:1996}. 
The quest was out to exploit this new capability in terms of  astrophysical processes. 
Already in early results, the characteristic signature of large-scale Galactic rotation was clearly indicated \citep{Diehl:2006d}, i.e., a blue shift when viewing towards the fourth quadrant (objects on Galactic orbits approaching) and a red shift when viewing towards the first quadrant (receding objects, on average).
With more exposure, this signature of large-scale Galactic rotation could be consolidated, as shown in Fig's~\ref{fig:al_longitudes} and \ref{fig:al_long-velocity}).
Thus, the observed gamma-ray flux could be translated into an observed total emitting mass of \Al, making use of geometrical models of how sources are distributed within the Galaxy, such as double-exponential disks and spiral-arm models \citep[see][for details]{Diehl:2006d}. 
The original result of 2.8$\pm$0.8~\Msol \citep{Diehl:2006d} was later revised and refined, as better geometrical models could be developed to account for foreground emission, and now suggests a somewhat lower total mass of Al in the Galaxy of between 1.8 ~\Msol and the earlier value \citep{Pleintinger:2020}. 

This observed mass of a single unstable isotope in the current Galaxy's interstellar medium can be converted into a rate of supernovae from massive stars (i.e., core-collapse supernovae). 
The supernova rate is key to driving turbulence within the interstellar medium \citep{Krumholz:2018,Koo:2020}, hence a key parameter to understand the dynamical state of interstellar medium.
Using \Al yields from theoretical models of massive stars and their supernovae, \citet{Diehl:2006d} derived a core-collapse supernova rate of 1.9$\pm$1.1 events per century for our Galaxy; this value was updated with better \Al mass measurements and model yields to 1.4$\pm$1.1~century$^{-1}$ \citep{Diehl:2018,Pleintinger:2020}, or one such supernova in our Galaxy every 71 years.
At face value, the 294 supernova remnants observed in the Galaxy \citep{Green:2019} thus present a tension with this value, as they would suggest a maximum sampling age of 21000~y, while significantly larger ages have been inferred for some of these. This, however, may be just another illustration of the dependence of the supernova remnant appearance on their surroundings.

The comparison of the observed velocity signature from large-scale Galactic rotation for \Al emission \citep{Kretschmer:2013} with the velocity signatures of other objects that orbit in such large-scale Galactic motion, however, lead to another puzzle (Fig.~\ref{fig:al_long-velocity}).
The underlying question already had been opened by the GRIS measurement \citep{Naya:1996}: 
 if interpreted as kinematic Doppler broadening of astrophysical origin, translates into an \Al motion of 540~km~s$^{-1}$. 
Considering the $1.04 \times 10^6$~y decay time of $^{26}$Al, such a large velocity observed for averaged interstellar decay of \Al would naively translate into kpc-sized cavities around \Al~ sources, so that velocities at the time of ejection would be maintained during the radioactive lifetime. An alternative hypothesis was that major fractions of  \Al would be condensed on grains, which would maintain ballistic trajectories in tenuous interstellar medium \citep{Chen:1997,Sturner:1999}. 

\begin{figure}[t]  
\centering 
  \includegraphics[width=0.45\linewidth]{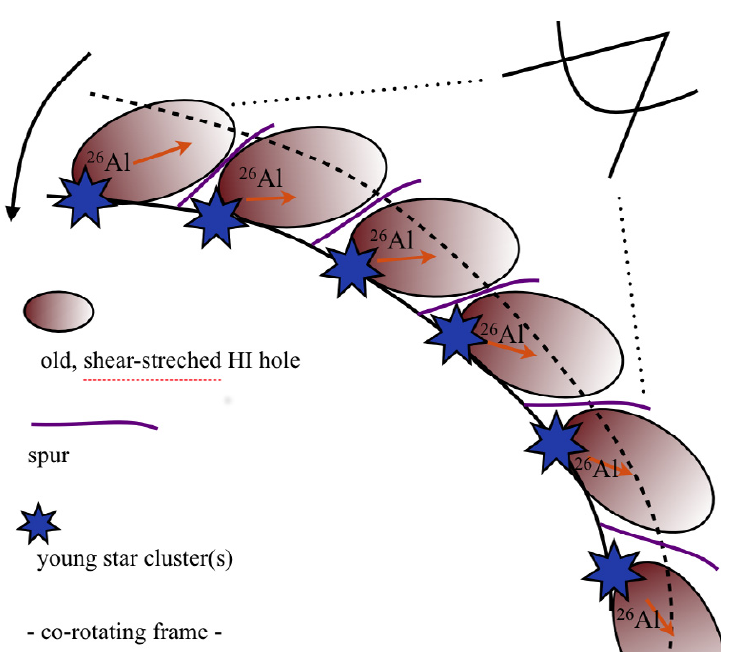}
\caption{A model for the different longitude-velocity signature of \Al, assuming \Al\ blown into inter-arm cavities at the leading side of spiral arms \citep{Krause:2015}. }
\label{fig:spiralarmbubbles} 
\end{figure}   

The velocities seen for \Al throughout the plane of the Galaxy with SPI on INTEGRAL  by \citet{Diehl:2006d} and \citet{Kretschmer:2013} exceed the velocities measured for several other objects such as molecular clouds, stars, and (most precisely measured velocities) maser sources by typically as much as 200~km~s$^{-1}$ (see Fig.~\ref{fig:al_long-velocity}). 
This high apparent velocity seen for bulk motion of decaying $^{26}$Al means that $^{26}$Al velocities remain higher than the velocities within typical interstellar gas for 10$^6$ years. Additionally, there is a bias for this excess average velocity in the direction of Galactic rotation. 
This has been interpreted as $^{26}$Al decay occurring preferentially within large cavities (superbubbles), which are elongated away from sources into the direction of large-scale Galactic rotation (Fig.~\ref{fig:spiralarmbubbles}). 
If such cavities are interpreted as resulting from the early onset of massive-star winds in massive-star groups, wind-blown bubbles would arise and characterise the source surroundings at times when stellar evolution terminates in the core-collapse of most-massive stars. Such  wind-blown superbubbles around massive-star groups plausibly extend further in space in forward directions and away from spiral arms (that host the sources), as has been seen in images of interstellar gas from other galaxies to occur \citep[][and references therein]{Schinnerer:2019}.  
Such superbubbles can extend up to kpc \citep{Krause:2015} \citep[see also][]{Rodgers-Lee:2019,Krause:2021}), which allows \Al to move at speeds of the sound velocity within superbubbles, and not much decelerated until its decay and gamma-ray emission.
One of the main open questions herein is the dynamics of superbubbles towards the Galactic halo above the disk; this is perpendicular to the line of sight, so that measurements of Doppler velocities cannot provide an answer.

The comparison of spatial distributions of candidate sources with the observed emission \citep{Diehl:1996b} suggested that the ejection of $^{26}$Al produced extended emission.
The inferred galactic emission scale height of about 180~pc \citep[see also][for improved constraints from \Al observations]{Wang:2009} clearly exceeds the scale height of parental molecular gas (50~pc), and is closer to the scale height inferred for OB associations in the nearby Galaxy \citep{Maiz-Apellaniz:2004}.  
In first attempts to understand this aspect, simulations can be exploited: In 3D hydro-simulations that provide results within the Galactic plane which compare well with observations, the simulation results towards other directions should be credible as well.
Recent years have provided computational power for large-scale simulations of this type \citep{Fujimoto:2018,Rodgers-Lee:2019}, dedicated to trace the fate of interstellar-medium evolution on a time scale of $\ge$10$^6$ years for major parts of the entire Galaxy. 
These simulations show consistently that chimneys reach out of the Galactic disk and towards the halo in regions of intense nucleosynthesis activity.
But, as a note of caution, \citet{Pleintinger:2019} have pointed out that such simulations cannot be interpreted literally in terms of spatial aspects, because initial conditions adopted herein do not properly represent the aspect of observing the Galaxy from our vantage point in the solar system.
Comparisons with expectations can be made in different ways. 
Biases in both the observations and the theoretical predictions require great care in drawing astrophysical conclusions \citep{Pleintinger:2019}. In particular, theoretical predictions often need to make assumptions about our Galaxy and its morphology. These are particularly critical for the vicinity of the solar position, as nearby sources would appear as bright emission that may dominate the signal \citep{Fujimoto:2020} .
As shown by  \citet{Pleintinger:2019}, generic characteristics of the interstellar medium morphology can well be extracted from comparisons of simulations to observations, but locations and extents of such chimneys require realism of the setup of simulations near the clusters of nucleosynthesis sources.
Therefore, it is necessary to develop an understanding of how such clusters of sources act to shape their environments.

\begin{figure}[t]  
\centering 
\includegraphics[width=0.9\columnwidth]{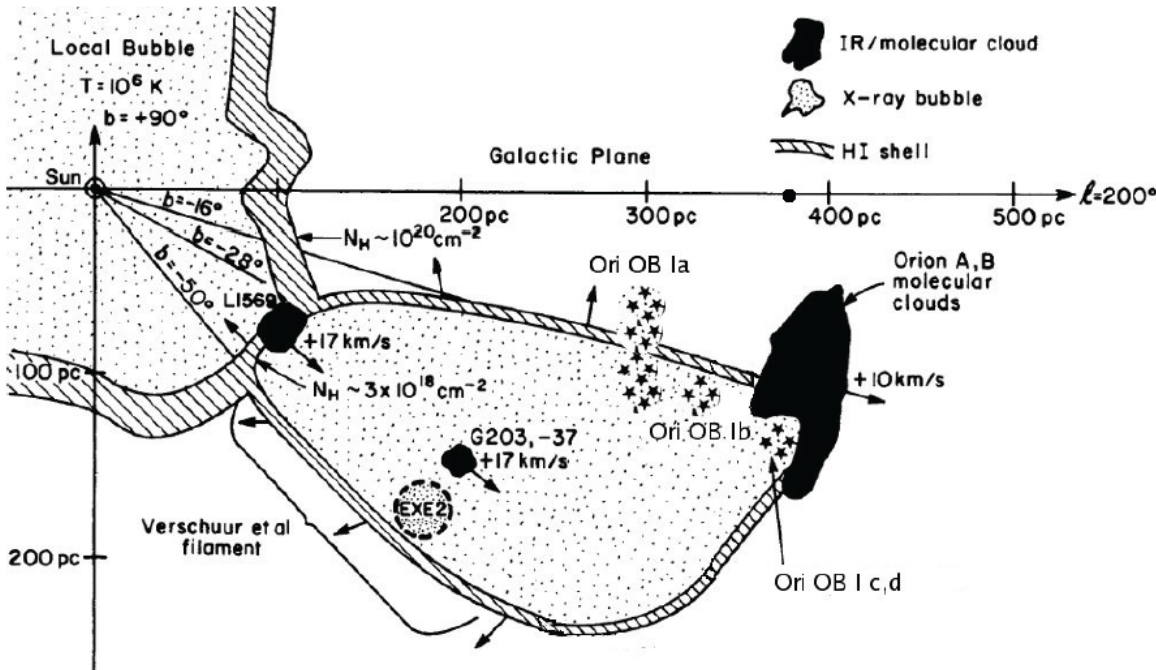}
\caption{The Orion region with the Eridanus cavity, illustrated from combined astronomical constraints \citep{Fierlinger:2012}. This region at a distance of about 450 pc for its massive stars of the Ori-Ob1 association and its subgroups forms an ideal laboratory to study superbubble astrophysics: The Eridanus cavity extends from the molecular clouds Orion A and B towards the Sun, and stars as well as gas structures are well mapped.}
\label{fig:ori-eri} 
\end{figure}   
 
\subsection{Interstellar environments of nucleosynthesis sources}\label{sec:astro-environments}  
The processes shaping interstellar environments of sources of nucleosynthesis are of interest, as they (i) determine the creation or operation of nucleosynthesis sources, and (ii) as they transport newly-produced nuclei from sources into the wider interstellar medium and towards places of interest (observers, newly-forming objects).
Both these aspects involve processes with characteristic time scales that are long compared to individual source/environment interactions (here, 10$^5$ years may be considered an upper bound, i.e., the oldest observable supernova remnant phenomena), and short compared to time scales of galactic evolution as a whole (here, 10$^8$ years may be considered a lower bound, i.e., time scales of a galaxy collision, or of spiral-arm formation or dissolving times, or the duration of an orbit of galactic rotation about its centre).
 
The analysis of \Al spectra from specific regions hosting clusters of nucleosynthesis sources was extended towards multi-messenger studies using the stellar census and information on atomic and molecular lines from radio data, as well as hot plasma from X-ray emission \citep{Voss:2009,Voss:2010a,Krause:2013,Krause:2014a}.
Such studies have taught us that the ejection of new nuclei and their feeding into next-generation stars apparently is a complex process, less simple than the \emph{instantaneous recycling} approximation in many chemical evolution considerations may assume. 
But being accessible to observational constraints such as from $\gamma$-ray line measurements shown for this case of \Al, the ejection and transport of new nuclei can now be studied in more detail.
\Al~ observations and their analyses have led to an \emph{\Al~ astronomy} within studies of stellar feedback and massive-star nucleosynthesis \citep[e.g.,][]{Krause:2020}. 

\begin{figure}[t]  
\centering 
\includegraphics[width=0.6\columnwidth]{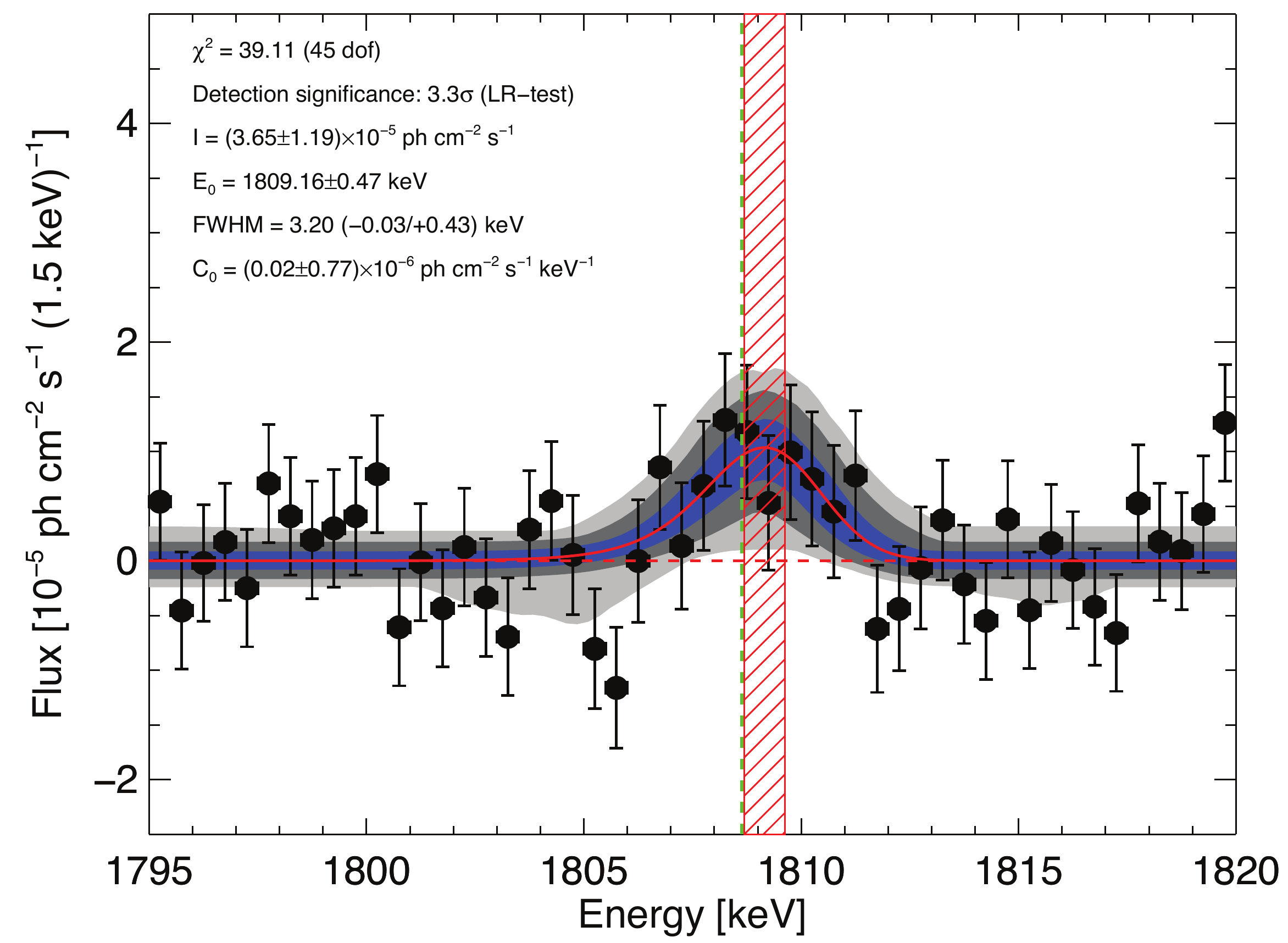}
\caption{ The \Al\ line measurement from the Orion region indicates the blue shift that would be expected from the geometry (Fig.~\ref{fig:ori-eri} with ejecta streaming towards us from the massive star groups located at the far side of the Eridanus cavity \citep{Siegert:2017}, although statistical precision is limited and deeper measurements are needed. }
\label{fig:al-Orion} 
\end{figure}

The Orion region has been understood as most-nearby example of massive star formation \citep{Bally:2008}. 
Star formation within the Orion A and B molecular clouds is understood to have created clusters of massive stars, the youngest now still observable as subgroups of the Orion OB1 association.
Fig.~\ref{fig:ori-eri} shows the geometry of this region, collected from various literature by \citet{Fierlinger:2012}, as seen from the Sun at a distance of about 450~pc. Particularly interesting is the cavity called Eridanus, which extends from the Orion molecular clouds towards the Sun, and is delineated in observations from atomic hydrogen (HI) as the cavity walls show enhanced gas density, as well as in X~rays from hot plasma as the cavity interior is heated by winds and supernovae from massive stars. 
\citet{Krause:2014a} have simulated a model for this setting of a cavity created by the energy (and mass) outputs from a cluster of stars. They show how overlapping of wind-blown bubbles leads to formation of a superbubble, and how later supernova explosions lead to periodic variations between excess heating and excess cooling in the interior of the cavity.
The deep observations with INTEGRAL/SPI later were able to resolve the \Al decay gamma-ray line, and show an indication of the Doppler blue-shift of the line, as ejecta from Orion OB1 stars expand into this cavity and towards the Sun (Fig.~\ref{fig:al-Orion}).

Population synthesis  had been employed as a tool to understand the collective action of a group of stars \citep{Leitherer:1999}. 
In our implementation of such an accumulation of mass and energy ejections as extracted from stellar-evolution modeling results \citep{Voss:2009}, we have shown how radioactive ejecta from \Al and \Fe can provide additional diagnostics of massive-star feedback.
The Orion region matches such mode expectations, within the uncertainties of stellar census and subgroup ages, of energy estimated from the Eridanus cavity extents, and of observed radioactive emission \citep{Voss:2010}.

As shown above (Fig.~\ref{fig:Al-map}), there are many (at least several hundreds) of such massive-star cluster regions within the Galaxy.
The Orion region obeys a favourable location for such studies, being sufficiently close for observational detail on stars and gas, yet being located at a distance away from the plane of the Galaxy, so that this region stands out from surrounding sky and can be distinguished from background at larger distances.
Difficulties from more-nearby aspects have become evident, as the stellar groups of the Scorpius-Centaurus associations at only 140~pc distance from the Sun have been studied \citep{Preibisch:2002,Poppel:2010}.
Nevertheless, also in this region, our general understanding of how massive stars form from molecular gas and erode their parental clouds through winds and supernovae is confirmed, when the current population of stars is compared against traces of past activity that pushes such gas and enriches it with radioactive nuclei \citep[see detailed discussion in][]{Krause:2018}.
In this nearby example, one recognises  the activity over time, and may describe it as \emph{surround and squish} action on parental molecular clouds, thus creating successive generations of massive cloud groups.
This is reminiscent, but still significantly different from the \emph{triggered star formation} scenario that \citep{Preibisch:2002} had proposed being exemplified by the Scorpius-Centaurus region's objects. 

Our solar system has been recognised also to be enclosed within a major cavity, called \emph{the Local Bubble} \citep{Breitschwerdt:1996}. 
It is interesting to study the origins of this cavity, and to see if this can be related to the drivers of interstellar-medium turbulence, as discussed above \citep{Lallement:2007}.
Radioactive admixtures to supernova ejecta again prove helpful in this context:
The analysis of ocean crust samples from the deep and remote Pacific ocean floor had revealed inclusions of radioactive $^{60}$Fe atoms \citep{Knie:1999,Knie:2004}. 
Through the extremely-sensitive \emph{accelerator mass spectrometry}  \citep{Kutschera:1990}, which allows to find rare atomic nuclear species with a concentration as low as one in 10$^{-15}$,  with 69 $^{60}$Fe nuclei identified in the first such quantitative analysis  \citep{Knie:1999}. By now, other ocean crust and sediment samples across the surface of Earth, and also lunar samples, all have obtained through the radioactive $^{60}$Fe nucleus the detections of such ashes attributed to core-collapse supernovae \citep{Wallner:2016}. 
Importantly, the sedimentation process includes also records of other radioactive nuclei that can serve as a sedimentation clock, such as $^{10}$Be produced in the Earth's atmosphere from cosmic-ray spallation. Using this to determine ages of depths within the samples, it is now clear that the $^{60}$Fe influx onto Earth is not homogeneous in time; rather, a clear peak of influx is found about 3 million years before present time, and a secondary peak is indicated near 8 Myrs \citep{Wallner:2016}. 
In an attempt to understand the astrophysical processes of nearby supernova explosions and ejecta propagation in turbulent and cavity-shaped interstellar medium, 3D studies of nearby objects have helped: It now appears that a nearby group of stars can be traced back in time to have been close to the Sun within the distance currently enclosed by the Local Bubble, and  its supernovae have been creating this local bubble \citep{Breitschwerdt:2006}. 
Moreover, state-of-the-art 3D simulations used to trace the co-evolution of interstellar morphology and transport of supernova ejecta \citep{Breitschwerdt:2016,Schulreich:2017} have shown that one or several supernovae originating from the same group could have shed their ejecta onto Earth.
Note that the peak in time with enhanced $^{60}$Fe influx appears too broad to be plausibly consistent with a single supernova ejecta cloud sweeping over the solar system  \citep{Breitschwerdt:2016}. 
These same simulations leave as another possibility the cavity walls of the local bubble, as it accumulates ejecta produced within the cavity: The Sun came close to the cavity walls at a time around 3~Myrs ago, and may thus have faced enhanced influx of rather recent supernova ejecta. 

It is interesting to note that recent analyses of cosmic-ray measurements with the CRIS instrument aboard the ACE satellite \citep{Binns:2016} have also revealed a signal from radioactive $^{60}$Fe. Unlike most other radioactive nuclei found in cosmic rays, $^{60}$Fe at the detected amount cannot be produced in high-energy reactions (spallation reactions) as cosmic rays hit interstellar gas.
Therefore, this current-day influx of $^{60}$Fe nuclei is yet another proof that supernova activity occurs within reach of cosmic-ray transport towards and into the solar system. 
Clearly, the ocean crust signal demonstrates a special enhancement in time, 3 to 8 Myrs ago; but the cosmic-ray measurement shows that even today the interstellar medium in the solar vicinity is shaped by supernova explosions. Note that \citet{Binns:2016} estimate a distance of the $^{60}$Fe sources of up to 1~kpc, from estimations of cosmic-ray transport.

With this experience in detection of radioactive traces of cosmic nucleosynthesis in terrestrial archives, there is another result that should be mentioned,
 in an attempt to study the origins of heaviest elements above the iron group. 
These are attributed to the neutron capture processes of the s and r~process \citep{Burbidge:1957}.
For a long time, core-collapse supernovae and their neutrino-driven winds have been considered the most-promising sources of r process nuclei \citep{Cowan:1982,Thielemann:1993}. 
But galactic archeology through determination of stellar surface abundances for stars of different ages (with Fe abundance as a proxy for age) had shown that the early Galaxy was not enriched with r-process elements (using Eu as a proxy) in the same way as expected from core-collapse supernovae, and traced in buildup of abundances of metals typical for these explosions, such as O, Mg, and Si. Fluctuations in time of Eu abundances suggested that it was not a frequently-occurring source such as common supernovae which would dominate r-process element production.
Therefore, rare but efficient alternative sources were studied and proposed, such as neutron star collisions and rare core-collapse supernovae with high characteristic magnetic fields. 
The detection of a kilonova coincident with the collision of two neutron stars as found in its characteristic gravitational-wave signal was a milestone for supporting such theoretical studies. 
Early enthusiasm led to claims that now the source of r-process elements had been found.
But closer inspection of expected rates and their evolution in time showed that also neutron star collisions cannot plausibly satisfy the constraints of galactic archeology, and we may face a superposition of contributions from more than one rare type of nucleosynthesis sources \citep{Cote:2019,Thielemann:2020}.

$^{244}$Pu has a radioactive lifetime of 1.2~10$^8$~y. A detection of $^{244}$Pu on Earth thus would represent a record of r-process nucleosynthesis near the solar system, similar to what has been found with $^{60}$Fe from core-collapse supernovae.
The ocean crust samples therefore have also been searched for rare nuclei at higher masses, which could shed light on r-process origins \citep{Wallner:2015}. 
Detecting merely one nucleus in an analysis of 11 ocean crust samples led \citet{Wallner:2015} to conclude that core-collapse supernovae of the frequently-occurring type cannot be responsible for r-process materials such as $^{244}$Pu, as a much higher detection count would be expected.
Of course, propagation of ejecta from a nucleosynthesis event into ocean floor sediments is a complex sequence of different astro-physical processes. 
But even accounting for all uncertainties involved, it is remarkable that detection of even individual and single radioactive nuclei in terrestrial material can provide astrophysical constraints on types of nucleosynthesis events.  

\subsection{Conclusions}\label{sec:conclusions}
 
The interstellar medium receives injections of turbulent energy and of matter from nucleosynthesis as massive star winds and supernova explosions occur. Within the nucleosynthesis products, radioactive isotopes are co-injected, and can be used to trace both the injection sources as well as the flow of ejecta materials. 
In the medium surrounding nucleosynthesis sources, short-lived radioisotopes such as $^{44}$Ti decay and reveal conditions within the individual sources. The Cas A supernova remnant is the most-prominent example of such events, and shows through its radioactive decay gamma rays that ejecta propagate in major clumps, rather than in concentric shells, into the interstellar medium.
More long-lived radioisotopes such as $^{26}$Al and $^{60}$Fe accumulate around sources, as their decay times are longer than the frequency of occurrence of source ejections in clusters of massive stars. Measuring the characteristic gamma rays from their decays have revealed that large cavities around such massive-star clusters are frequently found, and mediate the flow of nucleosynthesis ejecta towards dissolving later in the interstellar medium. 
Terrestrial and lunar sediments have also shown traces of cosmic  $^{60}$Fe, and thus witness such massive-star activity having occurred recently in the vicinity of the solar system. Recently, radioactive $^{244}$Pu has been measured in the same sediments. This exciting finding can be interpreted in terms of nearby supernova activity, and possibly as a constraint on r-process nucleosynthesis happening within some rare subset of these. 

These examples underline the complementary capabilities of interstellar medium studies through radioactive materials, and prominently with cosmic gamma ray spectrometers. 
INTEGRAL currently is the only space mission that features a spectrometer for nuclear cosmic gamma rays. This mission is far beyond its originally-planned lifetime of 5 years, and may not last for too long; it will end its orbital lifetime in 2029.
New missions towards these science goals and broader multi-messenger studies at up to two orders of magnitude better sensitivity have been proposed, such as the Astrogam mission \citep{de-Angelis:2018}. It remains to be seen if the astrophysical community can find budgets for such a mission, within the suite of exciting other bis projects of astronomy.

%
 \acknowledgments
This work was supported by the Max Planck Society, by DLR and ESA, and through EU COST Action 160117.


%
\bibliographystyle{spbasic}  

%

\end{document}